\newcommand{\xbS}{\Sigma}

\newcommand{\xba}{\alpha}

\newcommand{\xbe}{\in}

\newcommand{\xbm}{\mu}

\newcommand{\xbs}{\sigma}
\newcommand{\xbt}{\tau}

\newcommand{\xCN}{\neg}
\newcommand{\xCO}{ }
\newcommand{\xCQ}{\emptyset}

\newcommand{\xCc}{<}

\newcommand{\xCe}{>}

\newcommand{\xcA}{\forall}
\newcommand{\xcB}{\subsetneqq}
\newcommand{\xcC}{\not\subseteq}

\newcommand{\xcE}{\exists}

\newcommand{\xcS}{\bigcap}
\newcommand{\xcT}{\bot}

\newcommand{\xcV}{\bigcup}

\newcommand{\xcc}{\subseteq}

\newcommand{\xcg}{\geq}
\newcommand{\xch}{\Rightarrow}

\newcommand{\xcj}{\Leftrightarrow}
\newcommand{\xck}{\leq}

\newcommand{\xco}{\vee}
\newcommand{\xcp}{\rightarrow}

\newcommand{\xcs}{\cap}
\newcommand{\xct}{\top}
\newcommand{\xcu}{\wedge}
\newcommand{\xcv}{\cup}

\newcommand{\xcz}{\Box}

\newcommand{\xDH}{\item }

\newcommand{\xDK}{\otimes}

\newcommand{\xDN}{\ominus}

\newcommand{\xdx}{{\cal X}}

\newcommand{\xEI}{\begin{itemize}}
\newcommand{\xEJ}{\end{itemize}}

\newcommand{\xEd}{\neq}

\newcommand{\xEh}{\begin{enumerate}}

\newcommand{\xEj}{\end{enumerate}}

\newcommand{\xEn}{\begin{description}}
\newcommand{\xEp}{\end{description}}

\newcommand{\xew}{\sqcap}

\newcommand{\xfB}{\uparrow}

\newcommand{\xfw}{\sqcup}

\newcommand{\bl}{\begin{lemma} \rm}
\newcommand{\el}{\end{lemma}}
\newcommand{\br}{\begin{remark} \rm}
\newcommand{\er}{\end{remark}}
\newcommand{\be}{\begin{example} \rm}
\newcommand{\ee}{\end{example}}
\newcommand{\bco}{\begin{corollary} \rm}
\newcommand{\eco}{\end{corollary}}
\newcommand{\bc}{\begin{claim} \rm}
\newcommand{\ec}{\end{claim}}
\newcommand{\bfa}{\begin{fact} \rm}
\newcommand{\efa}{\end{fact}}
\newcommand{\bp}{\begin{proposition} \rm}
\newcommand{\ep}{\end{proposition}}
\newcommand{\bd}{\begin{definition} \rm}
\newcommand{\ed}{\end{definition}}
\newcommand{\bcs}{\begin{construction} \rm}
\newcommand{\ecs}{\end{construction}}
\newcommand{\bcd}{\begin{condition} \rm}
\newcommand{\ecd}{\end{condition}}
\newcommand{\bt}{\begin{theorem} \rm}
\newcommand{\et}{\end{theorem}}
\newcommand{\bn}{\begin{notation} \rm}
\newcommand{\en}{\end{notation}}
\newcommand{\bfi}{\begin{bild} \rm}
\newcommand{\efi}{\end{bild}}
\newcommand{\bsta}{\begin{statement} \rm}
\newcommand{\esta}{\end{statement}}
\newcommand{\bcom}{\begin{comment} \rm}
\newcommand{\ecom}{\end{comment}}
\newcommand{\bdia}{\begin{diagram} \rm}
\newcommand{\edia}{\end{diagram}}

\newcommand{\bfc}{\begin{figure}[htb] \begin{center}}
\newcommand{\efc}{\end{center} \end{figure}}

\documentclass{article}

\usepackage{amssymb,latexsym,epic,eepic,rotating}

\title{Operations on Partial Orders
\thanks{File: par
}
}

\author{Karl Schlechta
\thanks{
schcsg@gmail.com - https://sites.google.com/site/schlechtakarl/ -
Koppeweg 24, D-97833 Frammersbach, Germany}
\thanks{
Retired, formerly: Aix-Marseille Universit\'{e}, CNRS, LIF UMR 7279, F-13000
Marseille, France
}
}

\begin{document}

\newtheorem{lemma}{Lemma}[section]
\newtheorem{theorem}[lemma]{Theorem}
\newtheorem{proposition}[lemma]{Proposition}
\newtheorem{corollary}[lemma]{Corollary}
\newtheorem{claim}[lemma]{Claim}
\newtheorem{fact}[lemma]{Fact}
\newtheorem{remark}[lemma]{Remark}
\newtheorem{definition}{Definition}[section]
\newtheorem{construction}{Construction}[section]
\newtheorem{condition}{Condition}[section]
\newtheorem{example}{Example}[section]
\newtheorem{notation}{Notation}[section]
\newtheorem{bild}{Figure}[section]
\newtheorem{comment}{Comment}[section]
\newtheorem{statement}{Statement}[section]
\newtheorem{diagram}{Diagram}[section]

\renewcommand{\labelenumi}
  {(\arabic{enumi})}
\renewcommand{\labelenumii}
  {(\arabic{enumi}.\arabic{enumii})}
\renewcommand{\labelenumiii}
  {(\arabic{enumi}.\arabic{enumii}.\arabic{enumiii})}
\renewcommand{\labelenumiv}
  {(\arabic{enumi}.\arabic{enumii}.\arabic{enumiii}.\arabic{enumiv})}

\maketitle

\setcounter{secnumdepth}{3}
\setcounter{tocdepth}{3}

\begin{abstract}

We define analogues of Boolean operations on not necessarily complete partial
orders, they often have as results sets of elements rather than single elements.
It proves useful to add to such sets X if they are intended to be sup(X) or
inf(X), even if sup and inf do not always exist.

We then define the height of an element as the maximal length of chains going
from BOTTOM to that element, and use height to define probability measures.

\end{abstract}

\tableofcontents
\clearpage

%
%
%
\section{
Introduction
}

\label{Section Intro}
\subsection{
Overview
}

We define here Boolean operations on not necessarily complete partial
orders,
and then probability measures on such orders.

In a way, this is a continuation of work in
 \cite{Leh96} and  \cite{DR15}.

 \xEh
 \xDH Boolean operators

In the rest of
Section 
\ref{Section Intro} (page 
\pageref{Section Intro}), we discuss in general terms how to find and
judge a generalizing definition.

In Section 
\ref{Section Boole} (page 
\pageref{Section Boole}), we first give the basic definitions, see
Definition \ref{Definition Elements} (page \pageref{Definition Elements}),
they are quite standard, but due to incompleteness,
the results may be sets of several elements, and not single elements (or
singletons). This forces us to consider operators on sets of elements,
which sometimes complicates the picture, see
Definition \ref{Definition Sets} (page \pageref{Definition Sets}).
An alternative definition for sets is given in
Definition 
\ref{Definition Alternative} (page 
\pageref{Definition Alternative}), but
Fact \ref{Fact Alternative} (page \pageref{Fact Alternative})  shows
why we will not use this definition.

We then discuss basic properties of our definitions in
Fact \ref{Fact Rules} (page \pageref{Fact Rules}).

We conclude in Section \ref{Section Sign} (page \pageref{Section Sign})
by defining a - to our knowledge -
novel way to treat absent elements, by giving an indication of what is
really meant.

 \xDH Height, size, and probability

In Section 
\ref{Section Height} (page 
\pageref{Section Height}), we introduce the ``height'' of an
element, as the
maximal length of a chain from $ \xcT $ to that element, see
Definition \ref{Definition Height} (page \pageref{Definition Height}).

In Section 
\ref{Section Sequences} (page 
\pageref{Section Sequences}), we argue that the situation for
sequences of partial
orders may be more complicated than their product - basically as
a warning about perhaps unexpected problems.

Section 
\ref{Section Proba} (page 
\pageref{Section Proba})  introduces two notions of size for sets of
elements,
one, Definition \ref{Definition P} (page \pageref{Definition P}),
by the maximal height of its elements, the other,
Definition \ref{Definition Mu} (page \pageref{Definition Mu}), as the sum of
their heights. We also discuss some basic properties of these
definitions.

 \xEj

This is a rough draft, and mainly intended to present ideas.
\subsection{
Adequacy of a definition
}

We are not perfectly happy with our generalizations of the usual
operations
of $ \xew,$ $ \xfw,$ and $ \xDN $ to not necessarily complete partial
orders. We looked at a
few alternative definitions, but none is fully satisfactory.

There are a number of possible considerations when working on a new
definition, here a generalization of a standard definition:
 \xEI
 \xDH
Do we have a clear intuition?
 \xDH
Is there a desired behaviour?
 \xDH
Are there undesirable properties, like trivialisation in certain cases?
 \xDH
Can we describe it as an approximation to some ideal? Perhaps with some
natural distance?
 \xDH
How does the new definition behave for the original situation, here
complete
partial orders, etc.?
 \xEJ

In Section \ref{Section Sign} (page \pageref{Section Sign}),
we discuss in preliminary outline
a (new) approach, by adding supplementary information to the results of
the operations, which may help further processing. Thus, it may help
to improve definitions of the operators. The operators now do not only
work on elements or sets of elements, but also the additional information,
e.g., instead of considering $X \xew Y,$ we consider $inf(X) \xew inf(Y),$
$sup(X) \xew sup(Y),$ etc., where
``inf'' and ``sup'' is the supplementary information.
\subsection{
Motivation
}

In reasoning about complicated situations, e.g. in
legal reasoning, see for instance  \cite{Haa14},
the chapter on legal probabilism,
classical probability theory is often criticised for
imposing comparisons which seem arbitrary.
Our approach tries to counter such criticism by a more
flexible approach.
\clearpage
\section{
Boolean operations in partial orders
}

\label{Section Boole}

Assume a finite partial order $(\xdx,<)$ with TOP, $ \xct,$ and BOTTOM,
$ \xcT,$ and
$ \xcT < \xct,$ i.e. $ \xdx $ has at least two elements. $<$ is assumed
transitive.
We do not assume that the order is complete.

We will not always detail the order, so if we do not explicitly say that
$x<y$ or $y<x,$ we will assume that they are incomparable - with the
exception
$ \xcT <x< \xct $ for any $x,$ and transitivity is always assumed to hold.
\subsection{
Definitions
}

\bd

$\hspace{0.01em}$


\label{Definition MinMax}

 \xEh
 \xDH
For $x,y \xbe \xdx,$ set $x \xfB y$ iff $a \xck x$ and $a \xck y$ implies
$a= \xcT.$
 \xDH
For $X \xcc \xdx,$ define

$min(X)$ $:=$ $\{x \xbe X:$ $ \xCN \xcE x' \xbe X.x' <x\}$

$max(X)$ $:=$ $\{x \xbe X:$ $ \xCN \xcE x' \xbe X.x' >x\}$
 \xEj

\ed

\bfa

$\hspace{0.01em}$


\label{Fact Up}

$x \xfB y,$ $x' \xck x$ $ \xcp $ $x' \xfB y.$

(Trivial by transitivity.)

\efa

\bd

$\hspace{0.01em}$


\label{Definition Set-Smaller}

We define

 \xEh
 \xDH
$X^{y}:=\{x \xbe X.x \xck y\}$
 \xDH
$X \xck Y$ iff $ \xcA x \xbe X \xcE y \xbe Y.x \xck y$
 \xDH
$X<Y$ iff $X \xck Y$ $ \xcE y \xbe Y \xcA x \xbe X^{y}.x<y$
 \xEj

\ed

\br

$\hspace{0.01em}$


\label{Remark Set-Smaller}

 \xEh
 \xDH
$X \xck \{ \xct \}$ (trivial).
 \xDH
$X \xcc Y$ $ \xch $ $X \xck Y$ (trivial).
 \xDH
The alternative definition:

$X \xck_{1}Y$ iff $ \xcA y \xbe Y \xcE x \xbe X.x \xck y$

does not seem right, as the example $X:=\{a, \xct \},$ $Y:=\{b\},$ and
$a<b$ shows, as then
$X \xck_{1}Y.$
 \xEj

\er

We want to define analogues of the usual boolean operators, written
here $ \xew,$ $ \xfw,$ $ \xDN.$

We will see below that the result of a simple operation will not always
give a
simple result, i.e. an element or a singleton, but a set with
several elements as result.
Consequently, we will, in the general case,
have to define operations on sets of elements, not only on single
elements.
Note that we will often not distinguish between singletons and their
element,
what is meant will be clear from the context.

\bd

$\hspace{0.01em}$


\label{Definition Elements}

 \xEh
 \xDH
Let $x,y \xbe \xdx.$ The ususal $x \xew y$ might not exist, as the order
is not necessarily
complete. So, instead of a single ``best'' element, we might have only a set
of ``good'' elements.

Define
 \xEh
 \xDH
$x \xew y$ $:=$ $\{a \xbe \xdx:$ $a \xck x$ and $a \xck y\}$

This is not empty, as $ \xcT \xbe x \xew y.$

If $X \xcc \xdx $ is a set, we define

$ \xew X$ $:=$ $\{a \xbe \xdx:$ $a \xck x$ for all $x \xbe X\}.$

In particular, $x \xew y \xew z$ $:=$ $\{a \xbe \xdx:$ $a \xck x,$ $a
\xck y,$ $a \xck z\}.$
 \xDH
We may refine, and consider

$x \xew' y$ $:=$ $max(x \xew y)$

Usually, also $x \xew' y$ will contain more than one element.

We will consider in the next section a subset $x \xew'' y$ of $x \xew'
y,$
but $x \xew'' y$ may still contain several elements.
 \xEj

 \xDH
Consider now $ \xfw.$ The same remark as for $ \xew $ applies here, too.

Define
 \xEh
 \xDH
$x \xfw y$ $:=$ $\{a \xbe \xdx:$ $a \xcg x$ and $a \xcg y\}.$ Note that $
\xct \xbe x \xfw y.$

If $X \xcc \xdx $ is a set, we define

$ \xfw X$ $:=$ $\{a \xbe \xdx:$ $a \xcg x$ for all $x \xbe X\}.$

In particular, $x \xfw y \xfw z$ $:=$ $\{a \xbe \xdx:$ $a \xcg x,$ $a
\xcg y,$ $a \xcg z\}.$
 \xDH
Next, we define

$x \xfw' y$ $:=$ $min(x \xfw y)$

Again, we will also define some $x \xfw'' y \xcc x \xfw' y$ later.
 \xEj

 \xDH
Consider now $ \xDN.$

Define
 \xEh
 \xDH Unary $ \xDN $
 \xEh
 \xDH
$ \xDN x$ $:=$ $\{a \xbe \xdx:$ $a \xfB x\},$ note that $ \xcT \xbe \xDN
x.$

If $X \xcc \xdx $ is a set, we define

$ \xDN X$ $:=$ $\{a \xbe \xdx:$ $a \xfB x$ for all $x \xbe X\}$
 \xDH
Define

$ \xDN' x$ $:=$ $max(\xDN x)$

$ \xDN' X$ $:=$ $max(\xDN X)$

Again, we will also define some $ \xDN'' x \xcc \xDN' x$ later.
 \xEj

It is not really surprising that the seemingly intuitively correct
definition
for the set variant of $ \xDN $ behaves differently from that for $ \xew $
and $ \xfw,$
negation often does this. We will, however, discuss an alternative
definition in Definition 
\ref{Definition Alternative} (page 
\pageref{Definition Alternative}), (3),
and show that it seems inadequate in
Fact \ref{Fact Alternative} (page \pageref{Fact Alternative}), (3).

 \xDH Binary $ \xDN $
We may define $x-y$ either by $x \xew (\xDN y)$ or directly:
 \xEh
 \xDH
$x \xDN y$ $:=$ $\{a \xbe \xdx:$ $a \xck x$ and $a \xfB y\},$ note again
that $ \xcT \xbe x \xDN y,$

and
 \xDH
$x \xDN' y$ $:=$ $max(x \xDN y)$

 \xEj

For a comparison between direct and indirect definition, see
Fact \ref{Fact Rules} (page \pageref{Fact Rules}), (4.4).
 \xEj
 \xEj

\ed

We turn to the set operations, so assume $X,Y \xcc \xdx $ are sets of
elements, and we
define $X \xew Y,$ $X \xfw Y.$

The natural idea seems to consider all pairs $(x,y),$ $x \xbe X,$ $y \xbe
Y.$

\bd

$\hspace{0.01em}$


\label{Definition Sets}

We define the set operators:
 \xEh
 \xDH $ \xew $
 \xEh
 \xDH $ \xew $

$X \xew Y$ $:=$ $ \xcV \{x \xew y:$ $x \xbe X,$ $y \xbe Y\}$

 \xDH $ \xew' $

$X \xew' Y$ $:=$ $max(X \xew Y)$
 \xEj
 \xDH $ \xfw $
 \xEh
 \xDH $ \xfw $

$X \xfw Y$ $:=$ $ \xcV \{x \xfw y:$ $x \xbe X,$ $y \xbe Y\}$

 \xDH $ \xfw' $

$X \xfw' Y$ $:=$ $min(X \xfw Y)$
 \xEj
 \xDH $ \xDN $

$ \xDN X$ and $ \xDN' X$ were already defined. We do not define $X \xDN
Y,$ but see it as an
abbreviation for $X \xew (\xDN Y).$
 \xEj

\ed

See Definition 
\ref{Definition Alternative} (page 
\pageref{Definition Alternative})  and
Fact \ref{Fact Alternative} (page \pageref{Fact Alternative})
for an alternative definition for sets, and its discussion.
\subsection{
Properties
}

We now look at a list of properties,
for the element and the set versions.

\bfa

$\hspace{0.01em}$


\label{Fact Versions-1}

Consider $ \xdx:=\{ \xcT,a,b, \xct \}$ with $a \xfB b.$
We compare $ \xew $ with $ \xew',$ $ \xfw $ with $ \xfw',$ and $ \xDN
$ with $ \xDN'.$
 \xEh
 \xDH
$ \xct \xew a=\{x:x \xck a\}=\{ \xcT,a\},$ so $ \xct \xew a \xEd \{a\},$
but ``almost'', and $ \xct \xew' a=max(\xct \xew a)=\{a\}.$
 \xDH
$ \xcT \xfw a=\{x:x \xcg a\}=\{ \xct,a\},$ so $ \xcT \xfw a \xEd \{a\},$
but ``almost'', and $ \xcT \xfw' a=min(\xcT \xfw a)=\{a\}.$
 \xDH
$ \xDN a=\{ \xcT,b\},$ $ \xDN' a=\{b\},$ and by
Definition 
\ref{Definition Elements} (page 
\pageref{Definition Elements}), (3.1.1),
$ \xDN \xDN a=\{ \xcT,a\},$
and $ \xDN' \xDN' a=\{a\}.$

 \xDH

Consider $X=\{a,b\} \xcc \xdx.$

Then by Definition \ref{Definition Sets} (page \pageref{Definition Sets}),
(1), $ \xct \xew X=\{a, \xcT \} \xcv \{b, \xcT \}=\{a,b, \xcT \},$ and
$ \xct \xew' X=max(\xct \xew X)=X.$

 \xDH

Consider again $X=\{a,b\} \xcc \xdx.$

Then by Definition \ref{Definition Sets} (page \pageref{Definition Sets}), (2),
$ \xcT \xfw X=\{a, \xct \} \xcv \{b, \xct \}=\{a,b, \xct \},$ and
$ \xcT \xfw' X=min(\xcT \xfw X)=X.$
 \xEj
Thus, $ \xew',$ $ \xfw',$ $ \xDN' $ seem the better variants.

\efa

\bd

$\hspace{0.01em}$


\label{Definition Alternative}

We define alternative set operators, and argue in
Fact \ref{Fact Alternative} (page \pageref{Fact Alternative})
that they do not seem the right definitions.

 \xEh
 \xDH $ \xew_{1},$ $ \xew_{2}$
 \xEh
 \xDH
$X \xew_{1}Y$ $:=$ $ \xcS \{x \xew y:$ $x \xbe X,$ $y \xbe Y\},$
 \xDH
$X \xew_{2}Y$ $:=$ $ \xew (X \xcv Y)$
 \xEj
 \xDH $ \xfw_{1},$ $ \xfw_{2}$
 \xEh
 \xDH
$X \xfw_{1}Y$ $:=$ $ \xcS \{x \xfw y:$ $x \xbe X,$ $y \xbe Y\},$
 \xDH
$X \xfw_{2}Y$ $:=$ $ \xfw (X \xcv Y)$
 \xEj
 \xDH $ \xDN_{1}$

$ \xDN_{1}X$ $:=$ $\{a \xbe \xdx $: $a \xfB x$ for some $x \xbe X\}$
 \xEj

\ed

\bfa

$\hspace{0.01em}$


\label{Fact Alternative}

Consider $ \xdx:=\{ \xcT,a,b, \xct \}$ with $a \xfB b,$ and $X=\{a,b\}
\xcc \xdx.$

 \xEh
 \xDH $ \xew_{1},$ $ \xew_{2}$

 \xEh
 \xDH $ \xew_{1}:$

Then by Definition 
\ref{Definition Alternative} (page 
\pageref{Definition Alternative}), (1.1),
$ \xct \xew_{1}X=\{a, \xcT \} \xcs \{b, \xcT \}=\{ \xcT \}.$
 \xDH $ \xew_{2}:$

Then by Definition 
\ref{Definition Alternative} (page 
\pageref{Definition Alternative}), (1.2),
$ \xct \xew_{2}X= \xew \{a,b, \xct \}=\{ \xcT \}.$
 \xEj

 \xDH $ \xfw_{1},$ $ \xfw_{2}$

 \xEh
 \xDH $ \xfw_{1}:$

Then by Definition 
\ref{Definition Alternative} (page 
\pageref{Definition Alternative}), (2.1),
$ \xcT \xfw_{1}X=\{a, \xct \} \xcs \{b, \xct \}=\{ \xct \}.$
 \xDH $ \xfw_{2}:$

Then by Definition 
\ref{Definition Alternative} (page 
\pageref{Definition Alternative}), (2.2),
$ \xcT \xfw_{2}X= \xfw \{a,b, \xcT \}=\{ \xct \}.$
 \xEj

 \xDH $ \xDN_{1}$

 \xEh
 \xDH $ \xcT \xbe X$ $ \xch $ $ \xDN_{1}X= \xdx.$ (Trivial)

 \xDH In particular, $ \xDN_{1} \xdx = \xdx,$ which seems doubtful.

 \xDH $X \xcc \xDN_{1} \xDN_{1}X$

By $ \xcT \xbe \xDN_{1}X$ and the above, $ \xDN_{1} \xDN_{1}X= \xdx.$

 \xDH $ \xDN_{1} \xDN_{1}X \xcc X$ fails in general.

Consider $ \xdx $ $:=$ $\{ \xcT,$ a, $b,$ $c,$ $d,$ $ \xct \}$ with
$d<b,$ $d<c.$

Then $ \xDN_{1}\{b\}=\{ \xcT,a\},$ $ \xDN_{1}\{ \xcT,a\}= \xdx,$ so $
\xDN_{1} \xDN_{1}\{b\} \xcC \{b\}.$

 \xDH $ \xDN_{1}$ so defined is not antitone

Consider $ \xdx $ $:=$ $\{ \xcT, \xct \},$ $X:=\{ \xct \},$ $X':= \xdx
,$ then $ \xDN_{1}X=\{ \xcT \},$ $ \xDN_{1}X' =\{ \xcT, \xct \}.$
 \xEj
 \xEj

Thus, the variants in
Definition \ref{Definition Alternative} (page \pageref{Definition Alternative})
do not seem adequate.

\efa

\bfa

$\hspace{0.01em}$


\label{Fact Rules}

Commutativity of $ \xew $ and $ \xfw $ is trivial. We check simple cases
like
$ \xct \xew x,$ $ \xcT \xfw x,$ show that associativity holds, but
distributivity
fails. Concerning $ \xDN,$ we see that $ \xDN \xDN x$ is not
well-behaved, and neither
is the combination of $ \xDN $ with $ \xfw.$

 \xEh
 \xDH $ \xew $ and $ \xew' $
 \xEh
 \xDH $ \xct \xew x=x$?

$ \xct \xew x$ $=$ $\{a \xbe \xdx:$ $a \xck x\}.$

$ \xct \xew' x$ $=$ $\{x\}.$
 \xDH $ \xct \xew X=X$?

$ \xct \xew X$ $=$ $\{a:$ $a \xck x$ for some $x \xbe X\}$

$ \xct \xew' X$ $=$ $max(X)$ - which is not necessarily $X$ (if there are
$a,a' \xbe X$ with
$a<a').$

 \xDH $x \xew x=x$?

$x \xew x$ $=$ $\{a \xbe \xdx:$ $a \xck x\}$

$x \xew' x$ $=$ $\{x\}.$
 \xDH $X \xew X=X$?

$X \xew X$ $=$ $ \xcV \{x \xew y:$ $x,y \xbe X\}.$

Note that $x \xew y \xcc x \xew x$ for all $x,y \xbe \xdx,$ thus
$X \xew X$ $=$ $ \xcV \{x \xew x:x \xbe X\}$ $=$ $\{a \xbe \xdx:$ $a \xck
x$ for some $x \xbe X\}.$

$X \xew' X$ $=$ $max(X)$ - which is not necessarily $X.$

 \xDH $x \xew y \xew z$ $=$ $x \xew (y \xew z)$?

Let $A:=$ $x \xew y \xew z$ $=$ $\{a:$ $a \xck x,$ $a \xck y,$ $a \xck
z\}$

Set $B:=$ $y \xew z$ $=$ $\{b:$ $b \xck y,$ $b \xck z\}$
 \xEh
 \xDH $ \xew $

We have to show $A=x \xew B.$

$x \xew B$ $=$ $ \xcV \{x \xew b:$ $b \xbe B\}$ by
Definition \ref{Definition Sets} (page \pageref{Definition Sets}), (1).

If $a \xbe A,$ then $a \xbe B,$ moreover $a \xck x,$ so $a \xbe x \xew a
\xcc x \xew B.$

Let $a \xbe x \xew B,$ then there is $b \xbe B,$ $a \xbe x \xew b.$
As $b \xbe B,$ $b \xck y,$ $b \xck z,$ so $a \xck x,$ $a \xck b \xck y,$
$a \xck b \xck z,$ so $a \xbe A$
by transitivity.

 \xDH $ \xew' $

Set $A':=max(A),$ $B':=max(B).$ $x \xew' B' $ $=$ $max(\xcV \{x \xew'
b:$ $b \xbe B' \}).$

Let $a \xbe A' \xcc A \xcc B,$ so there is $b' \xcg a,$ $b' \xbe B',$ and
by $a \xbe A,$ $a \xck x,$ so $a \xbe x \xew b'.$

Suppose there is $a' \xbe \xcV \{x \xew' b:$ $b \xbe B' \}$ $ \xcc $ $x
\xew B,$ $a' >a.$
Then by (1.5.1) $a' \xbe A,$ contradicting maximality of a.

Conversely,
let $a \xbe max(\xcV \{x \xew' b:$ $b \xbe B' \})$ $ \xcc $ $x \xew B,$
then $a \xbe A$ by (1.5.1). Suppose there is
$a' >a,$ $a' \xbe x \xew' y \xew' z,$ so we may assume $a' \xbe A',$
then
$a' \xbe max(\xcV \{x \xew' b:$ $b \xbe B' \}),$ as we just saw,
contradiction.

Thus, it works for $ \xew',$ too.

 \xEj
 \xEj

 \xDH $ \xfw $ and $ \xfw' $
 \xEh
 \xDH $ \xcT \xfw x=x$?

$ \xcT \xfw x$ $=$ $\{a \xbe \xdx:$ $a \xcg x\}.$

$ \xcT \xfw' x$ $=$ $\{x\}.$

 \xDH $ \xcT \xfw X=X$?

$ \xcT \xfw X$ $=$ $\{a:$ $a \xcg x$ for some $x \xbe X\}$

$ \xcT \xfw' X$ $=$ $min(X)$ - which is not necessarily $X.$

 \xDH $x \xfw x=x$?

$x \xfw x$ $=$ $\{a \xbe \xdx:$ $a \xcg x\}.$

$x \xfw' x$ $=$ $\{x\}.$

 \xDH $X \xfw X=X$?

$X \xfw X$ $=$ $ \xcV \{x \xfw y:$ $x,y \xbe X\}.$

Note that $x \xfw y \xcc x \xfw x$ for all $x,y \xbe \xdx,$ thus
$X \xfw X$ $=$ $ \xcV \{x \xfw x:x \xbe X\}$ $=$ $\{a \xbe \xdx:$ $a \xcg
x$ for some $x \xbe X\}.$

$X \xfw' X$ $=$ $min(X)$ - which is not necessarily $X.$

 \xDH $x \xfw y \xfw z$ $=$ $x \xfw (y \xfw z)$?

Let $A:=$ $x \xfw y \xfw z$ $=$ $\{a:$ $a \xcg x,$ $a \xcg y,$ $a \xcg
z\}$

Set $B:=$ $y \xfw z$ $=$ $\{b:$ $b \xcg y,$ $b \xcg z\}$
 \xEh
 \xDH $ \xfw $

We have to show $A=x \xfw B.$

$x \xfw B$ $=$ $ \xcV \{x \xfw b:$ $b \xbe B\}$ by
Definition \ref{Definition Sets} (page \pageref{Definition Sets}), (2).

If $a \xbe A,$ then $a \xbe B,$ moreover $a \xcg x,$ so $a \xbe x \xfw a
\xcc x \xfw B.$

Let $a \xbe x \xfw B,$ then there is $b \xbe B,$ $a \xbe x \xfw b.$
As $b \xbe B,$ $b \xcg y,$ $b \xcg z,$ so $a \xcg x,$ $a \xcg b \xcg y,$
$a \xcg b \xcg z,$ so $a \xbe A$
by transitivity.

 \xDH $ \xfw' $

Set $A':=min(A),$ $B':=min(B).$ $x \xfw' B' $ $=$ $min(\xcV \{x \xfw'
b:$ $b \xbe B' \}).$

Let $a \xbe A' \xcc A \xcc B,$ so there is $b' \xck a,$ $b' \xbe B',$ and
by $a \xbe A,$ $a \xcg x,$ so $a \xbe x \xfw b'.$

Suppose there is $a' \xbe \xcV \{x \xfw' b:$ $b \xbe B' \}$ $ \xcc $ $x
\xfw B,$ $a' <a.$
Then by (2.5.1) $a' \xbe A,$ contradicting minimality of a.

Conversely,
let $a \xbe min(\xcV \{x \xfw' b:$ $b \xbe B' \})$ $ \xcc $ $x \xfw B,$
then $a \xbe A$ by (2.5.1). Suppose there is
$a' <a,$ $a' \xbe x \xfw' y \xfw' z,$ so we may assume $a' \xbe A',$
then
$a' \xbe min(\xcV \{x \xfw' b:$ $b \xbe B' \}),$ as we just saw,
contradiction.

Thus, it works for $ \xfw',$ too.

 \xEj

 \xEj

 \xDH Distributivity for $ \xew,$ $ \xfw,$ $ \xew',$ $ \xfw' $
 \xEh
 \xDH $x \xew (y \xfw z)$ $=$ $(x \xew y) \xfw (x \xew z)$?

Let $ \xdx:=\{ \xcT,x,y,z, \xct \}.$

Then $y \xfw z=\{ \xct \},$ so $x \xew (y \xfw z)=\{x, \xcT \}.$

$y \xfw' z=\{ \xct \},$ so $x \xew' (y \xfw' z)=max(\{x, \xcT
\})=\{x\}.$

$x \xew y$ $=$ $x \xew z$ $=$ $\{ \xcT \},$ so $(x \xew y) \xfw (x \xew
z)$ $=$ $ \xdx.$

$x \xew' y$ $=$ $x \xew' z$ $=$ $\{ \xcT \},$ so $(x \xew' y) \xfw' (x
\xew' z)$ $=$ $min(\xdx)$ $=$ $\{ \xcT \}.$

So distributivity fails for both versions.

 \xDH $x \xfw (y \xew z)$ $=$ $(x \xfw y) \xew (x \xfw z)$?

Let $ \xdx:=\{ \xcT,x,y,z, \xct \}.$

$y \xew z$ $=$ $\{ \xcT \}$ $=$ $y \xew' z.$ $x \xfw y$ $=$ $\{ \xct \}$
$=$ $x \xfw' y,$ $x \xfw z$ $=$ $\{ \xct \}$ $=$ $x \xfw' z.$

$x \xfw \xcT $ $=$ $\{x, \xct \},$ $x \xfw' \xcT $ $=$ $\{x\}.$

$ \xct \xew \xct $ $=$ $ \xdx,$ $ \xct \xew' \xct $ $=$ $\{ \xct \}.$

So it fails again for both versions.
 \xEj

 \xDH $ \xDN $ and $ \xDN' $
 \xEh
 \xDH $ \xDN \xct $ $=$ $ \xcT $?

$ \xDN \xct $ $=$ $\{ \xcT \}$

$ \xDN' \xct $ $=$ $\{ \xcT \}$
 \xDH $ \xDN \xcT $ $=$ $ \xct $?

$ \xDN \xcT $ $=$ $ \xdx $

$ \xDN' \xcT $ $=$ $\{ \xct \}$

 \xDH $ \xDN \xDN x$ $=$ $x$?

Consider $ \xdx:=\{ \xcT,x',x,y, \xct \}$ with $x<x'.$

Then $ \xDN x=\{ \xcT,y\},$ $ \xDN' x=\{y\},$
$ \xDN (\xDN x)=\{ \xcT,x',x\},$
$ \xDN (\xDN' x)=\{ \xcT,x',x\},$
$ \xDN' (\xDN' x)=\{x' \},$
so it fails for both versions.

 \xDH $x \xew (\xDN y)$ $=$ $x \xDN y$?

$x \xDN y$ $=$ $\{a \xbe \xdx:$ $a \xck x$ and $a \xfB y\}.$

$ \xDN y$ $=$ $\{a \xbe \xdx:$ $a \xfB y\}.$

$x \xew (\xDN y)$ $=$ $ \xcV \{x \xew a:$ $a \xbe \xdx,$ $a \xfB y\}$
$=$ $\{b \xbe \xdx:$ $b \xck x$ and $b \xck a$ for some $a \xbe \xdx,$
$a \xfB y\}$ $=$
$\{b \xbe \xdx:$ $b \xck x$ and $b \xfB y\}$ by
Fact \ref{Fact Up} (page \pageref{Fact Up}).
 \xDH $x \xew (\xDN x)$ $=$ $ \xcT $?

$x \xew (\xDN x)$ $:=$ $ \xcV \{x \xew y:$ $y \xbe (\xDN x)\}$ $=$ $
\xcV \{x \xew y:$ $y \xfB x\}$ by $y \xbe (\xDN x)$ $: \xcj $ $y \xfB x.$
Let $a \xbe x \xew y$ for $y \xfB x,$ then $a \xck x$ and $a \xck y,$ so
$a= \xcT.$

 \xDH $X \xew (\xDN X)$ $=$ $ \xcT $?

$X \xew (\xDN X)$ $:=$ $ \xcV \{x \xew y:$ $x \xbe X,$ $y \xbe (\xDN
X)\}$ $=$ $ \xcV \{x \xew y:$ $x \xbe X,$ $y \xfB x' $ for all $x' \xbe
X\}$
by $y \xbe (\xDN X)$ $: \xcj $ $y \xfB x' $ for all $x' \xbe X.$
Conclude as for (4.5).

 \xDH $x \xDN x$ $=$ $ \xcT $?

$x \xDN x$ $=$ $\{a \xbe \xdx:$ $a \xck x$ and $a \xfB x\}$ $=\{ \xcT \}$
$=$ $x \xDN' x.$
 \xDH $x \xfw (\xDN x)$ $=$ $ \xct $?

Consider $ \xdx:=(\xcT,a,b,c,ab, \xct \},$ with $a<ab,$ $b<ab.$

Then $ \xDN a=\{b,c, \xcT \},$ and $a \xfw (\xDN a)$ $=$ $ \xcV \{a \xfw
b,$ $a \xfw c,$ $a \xfw \xcT \}$ $=$
$\{ab, \xct,a\}$ $ \xEd $ $\{ \xct \}.$

$ \xDN' a$ $=$ $\{b,c\},$ $a \xfw' (\xDN' a)$ $=$ $min(\{ab, \xct \}
\xcv \xct \})$ $=$ $\{ab\}$ $ \xEd $ $\{ \xct \},$
so it fails for both versions.

 \xDH $ \xDN $ is antitone: $X \xcc X' $ $ \xch $ $ \xDN X' \xcc \xDN X$

$a \xbe \xDN X' $ $ \xch $ $a \xfB x$ for all $x \xbe X',$ so $a \xfB x$
for all $x \xbe X' $ $ \xch $ $a \xbe \xDN X.$

 \xDH $X \xcc \xDN \xDN X$

$ \xDN X$ $:=$ $\{a:$ $a \xfB x$ for all $x \xbe X\}.$

Let $x \xbe X,$ $a \xbe \xDN X.$ By $a \xbe \xDN X,$ $x \xfB a,$ so $x
\xbe \xDN \xDN X.$

 \xDH $ \xDN \xDN X \xcc X$ fails in general.

Consider $ \xdx $ $:=$ $\{ \xcT,$ a, $b,$ $c,$ $d,$ $ \xct \}$ with
$d<b,$ $d<c.$

Then $ \xDN \{b\}=\{ \xcT,a\},$ $ \xDN \{ \xcT,a\}=\{b,c,d\},$ so $ \xDN
\xDN \{b\} \xcC \{b\}.$

 \xEj

 \xEj
\subsection{
Sets with a sign
}

\label{Section Sign}

\efa

We will outline here a - to our knowledge, new - approach, and code the
last operation into the result, so the ``same'' result of two different
operations will look differently, and the difference will be felt in
further processing the result,
see the following Example \ref{Example SupInf} (page \pageref{Example SupInf}).
See also Definition \ref{Definition P} (page \pageref{Definition P}).

Basically, we give not only the result, as well as we can, but also an
indication, what the intended result is, ``what is really meant'', the ideal
- even if we are unable to
formulate it, for lack of an suitable element.

More precisely, if the result is a set $X,$ but what we really want is
$sup(X),$
which does not exist in the structure $ \xdx,$ we will have the result
with the sign
sup, i.e., $sup(X),$ likewise inf and $inf(X),$
and further processing may take this into consideration.

In a way, it is a compromise. The full information gives all arguments and
operators, the basic information gives just the result, we give the result
with an indication how to read it.

\be

$\hspace{0.01em}$


\label{Example SupInf}

Let $ \xdx:=\{a,b,c,d,x,x',y,e,e',f,f' \}$ with

$e<c<x<a<f,$

$e<d<x' <b<f,$

$c<x' <a,$

$d<x<b,$

$e<y<f,$

$e' <x' <f',$

$e' <y<f',$

see Diagram \ref{Diagram Inf/Sup} (page \pageref{Diagram Inf/Sup}).
(The relations involving $ \xcT $ and $ \xct $ are not shown in the
diagram,
$ \xcT \xEd e,$ $ \xct \xEd f.)$

Then:
 \xEh
 \xDH
$a \xew b$ $=$ $\{x,x' \},$ more precisely $a \xew b$ $=$ $sup\{y:$ $y
\xck a,$ $y \xck b\}$ $=$ $sup\{x,x' \}$
- which does not exist, but we do as if, i.e., we give a ``label'' to
$\{x,x' \}.$

Reason:

We have $x<a,b,$ $x' <a,b,$ $sup\{x,x' \}$ is the smallest $z$ such that
$z>x,$ $z>x',$ thus
$z<a,$ $z<b,$ but this $z$ does not exist.
 \xDH
$a \xfw b$ $=$ $\{x,x' \},$ more precisely $a \xfw b$ $=$ $inf\{y:$ $y
\xcg a,$ $y \xcg b\}$ $=$ $inf\{x,x' \},$
- which does not exist, but we do as if, i.e., we give a ``label'' to
$\{x,x' \}.$

Reason:

We have $x>c,d,$ $x' >c,d,$ $inf\{x,x' \}$ is the biggest $z$ such that
$z<x,$ $z<x',$ thus
$z>c,$ $z>d,$ but this $z$ does not exist.

 \xDH
$ \xDN a=sup\{x:$ $x \xfB a\}.$

 \xEj

To summarize, we have $x,x' \xck sup\{x,x' \} \xck a,b$ and $c,d \xck
inf\{x,x' \} \xck x,x' $ - but
$inf\{x,x' \}$ and $sup\{x,x' \}$ need not exist.

\ee

Consider now $y \xew \{x,x' \},$ $y \xfw \{x,x' \},$ $ \xfB \{x,x' \},$
and $ \xDN \{x,x' \}$ to see the difference:
 \xEh
 \xDH $ \xew $
 \xEh
 \xDH
$y \xew sup\{x,x' \}$ $=$ $\{z:$ $z<y$ $ \xcu $ $(z<x$ $ \xco $ $z<x')\}$
$=$ $\{e',e\}$
 \xDH
$y \xew inf\{x,x' \}$ $=$ $\{z:$ $z<y$ $ \xcu $ $(z<x$ $ \xcu $ $z<x')\}$
$=$ $\{e\}$
 \xEj
 \xDH $ \xfw $
 \xEh
 \xDH
$y \xfw sup\{x,x' \}$ $=$ $\{z:$ $z>y$ $ \xcu $ $(z>x$ $ \xcu $ $z>x')\}$
$=$ $\{f\}$
 \xDH
$y \xfw inf\{x,x' \}$ $=$ $\{z:$ $z>y$ $ \xcu $ $(z>x$ $ \xco $ $z>x')\}$
$=$ $\{f',f\}$
 \xEj
 \xDH $ \xfB $
 \xEh
 \xDH
$ \xfB sup\{x,x' \}$ $=$ $\{a:$ $a \xfB x$ $ \xcu $ $a \xfB x' \}$ $=$ $\{
\xcT \}$
 \xDH
$ \xfB inf\{x,x' \}$ $=$ $\{a:$ $a \xfB x$ $ \xco $ $a \xfB x' \}$ $=$ $\{
\xcT,$ $e' \}$
 \xEj
 \xDH $ \xDN $
 \xEh
 \xDH
$ \xDN sup\{x,x' \}$ $=$ $\{a:$ $a \xfB x$ $ \xcu $ $a \xfB x' \}$ $=$ $\{
\xcT \}$
 \xDH
$ \xDN inf\{x,x' \}$ $=$ $\{a:$ $a \xfB x$ $ \xco $ $a \xfB x' \}$ $=$ $\{
\xcT,$ $e' \}$
 \xEj
 \xEj

Basically, we remember the last operation resulting in an intermediate
result,
but even this is not always sufficient as the example in
Fact 
\ref{Fact Rules} (page 
\pageref{Fact Rules}), (3.1), failure of distributivity, shows:
The intermediate results $y \xfw z,$ $x \xew y,$ $x \xew z$ are
singletons, so our idea
has no influence.

One could try to write everything down without intermediate results, but
one has to find a compromise between correctness and simplicity.

\clearpage

\begin{diagram}

\label{Diagram Inf/Sup}
\index{Diagram Inf/Sup}

\unitlength1.0mm
\begin{picture}(150,180)(0,0)

\put(0,175){{\rm\bf Diagram Inf/Sup }}
\put(60,175){Recall that $inf\{x,x'\}$ and $sup\{x,x'\}$ do not exist}

\put(40,2){e}
\put(40,5){\circle*{1}}
\put(40,5){\line(1,1){25}}
\put(40,5){\line(-1,1){25}}
\put(40,5){\line(2,1){50}}
\put(10,30){c}
\put(15,30){\circle*{1}}
\put(15,30){\line(0,1){100}}
\put(15,30){\line(1,1){75}}
\put(67,30){d}
\put(65,30){\circle*{1}}
\put(65,30){\line(0,1){100}}
\put(65,30){\line(-1,1){50}}

\put(40,52){$ inf\{x,x'\}=c \xfw d $}
\put(40,55){\circle*{1}}
\put(12,78){x}
\put(15,80){\circle*{1}}
\put(15,80){\line(1,1){50}}
\put(65,78){x'}
\put(65,80){\circle*{1}}

\put(90,52){e'}
\put(90,55){\circle*{1}}
\put(90,55){\line(-1,1){75}}
\put(90,55){\line(1,1){25}}

\put(40,102){$ sup\{x,x'\}=a \xfw b $}
\put(40,105){\circle*{1}}
\put(12,130){a}
\put(15,130){\circle*{1}}
\put(15,130){\line(1,1){25}}
\put(67,130){b}
\put(65,130){\circle*{1}}
\put(65,130){\line(-1,1){25}}

\put(117,80){y}
\put(115,80){\circle*{1}}
\put(115,80){\line(-1,1){25}}
\put(115,80){\line(-1,-2){25}}
\put(115,80){\line(-1,2){25}}
\put(90,107){f'}
\put(90,105){\circle*{1}}

\put(40,157){f}
\put(40,155){\circle*{1}}
\put(40,155){\line(2,-1){50}}

\put(40,-10){{\bf \Large $\xcT$}}
\put(40,165){{\bf \Large $\xct$}}

\end{picture}

\end{diagram}

\vspace{4mm}

\clearpage

\clearpage
\section{
Height
}

\label{Section Height}
\subsection{
Basic definitions
}

\bd

$\hspace{0.01em}$


\label{Definition Height}

Let $x \xbe \xdx.$

Set $ht(x):=$ the length of the longest chain from $ \xcT $ to $x$ - where
we count
the number of $<$ in the chain.

Let $X \xcc \xdx.$

Define

$maxht(X)$ $:=$ $\{x \xbe X:$ $ \xcA x' \xbe X.ht(x) \xcg ht(x')\}$ and

$minht(X)$ $:=$ $\{x \xbe X:$ $ \xcA x' \xbe X.ht(x) \xck ht(x')\}.$

\ed

In Section 
\ref{Section P'} (page 
\pageref{Section P'}), we give an alternative definition of a
probability
using height.

\bfa

$\hspace{0.01em}$


\label{Fact Height}

 \xEh
 \xDH
$ht(\xcT)=0,$ $ht(\xct)>0.$
 \xDH
$ht(x) \xck ht(\xct)$ for all $x \xbe \xdx.$
 \xDH
We have $x<y$ $ \xcp $ $ht(x)<ht(y)$ for all $x,y \xbe \xdx.$
 \xDH
If $x$ and $y$ are $<$-incomparable, it does not necessarily follow that
$ht(x)=ht(y).$

(This is trivial, as seen e.g. in the example $ \xdx:=\{ \xcT,a,a',b,
\xct \}$ with
$a<a',$ so $ht(a')=2,$ $ht(b)=1,$ and $a',b$ are incomparable.)
 \xDH
$maxht(X)$ $=$ $maxht(max(X)),$ $minht(X)$ $=$ $minht(min(X))$

 \xEj

\efa

\bd

$\hspace{0.01em}$


\label{Definition ht-Versions}

 \xEh
 \xDH
$x \xew'' y:=maxht(x \xew y),$

$X \xew'' Y:=maxht(X \xew Y).$
 \xDH
$x \xfw'' y:=minht(x \xfw y),$

$X \xfw'' Y:=minht(X \xfw Y).$
 \xDH
$ \xDN'' x:=maxht(\xDN x),$

$ \xDN'' X:=maxht(\xDN X),$

$x \xDN'' y:=maxht(x \xDN y).$

 \xEj
We might also have chosen $x \xew' y$ instead of $x \xew y,$
etc., by Fact \ref{Fact Height} (page \pageref{Fact Height}), (5).

\ed

\be

$\hspace{0.01em}$


\label{Example ht-Versions}

Consider $ \xdx:=\{ \xcT,a,b,b',c, \xct \},$ with $b<b'.$

Then $ \xDN c=\{ \xcT,a,b,b' \},$ $ \xDN' c=\{a,b' \},$ and $ \xDN''
c=\{b' \},$ so we lose important
information, in particular, if we want to continue with Boolean
operations.

For this reason, the versions $ \xew'',$ $ \xfw'',$ $ \xDN'' $ should
be used with caution.
\subsection{
Sequences
}

\label{Section Sequences}

\ee

\be

$\hspace{0.01em}$


\label{Example Seq}

In the second example, we compensate a loss in the second coordinate
by a bigger gain in the first. Thus, the situation in the product might
be more complex that the combined situations of the elements of the
product.
 \xEh
 \xDH
Consider
$ \xdx $ $:=$ $\{0,1\}$ and $ \xdx' $ $:=$ $\{0',1' \}$ with the natural
orders. In $ \xdx,$ $ht(1)=1,$
in $ \xdx',$ $ht(1')=1.$

Order the sequences in $ \xdx \xDK \xdx' $ by the value of the sequences,
defined as
their sum.

So in $ \xdx \xDK \xdx',$ $ \xcT =(0,0')<(0,1')<(1,1')= \xct,$
$(0,0')<(1,0')<(1,1'),$
and $ht((1,1'))=2.$
 \xDH
Consider now
$ \xdx $ $:=$ $\{0,2\}$ and $ \xdx' $ $:=$ $\{0',1' \}$ with the natural
orders. In $ \xdx,$ $ht(2)=1,$
in $ \xdx',$ $ht(1')=1$ again.

Order the sequences in $ \xdx \xDK \xdx' $ again by the value of the
sequences, defined as
their sum.

So in $ \xdx \xDK \xdx',$ $ \xcT $ $=$ $(0,0')$ $<$ $(0,1')$ $<$
$(2,0')$ $<$ $(2,1')$ $=$ $ \xct,$ and $ht((2,1'))=3.$
 \xEj

\ee

Of course, we use here additional structure of the components, sum and
difference.

In general, we may consider rules like:

$ \xbs \xbs' < \xbt \xbt' $ iff $ \xbs' \xbs < \xbt' \xbt $ and

$ \xbs \xbs' < \xbt \xbt' $ iff $ \xbs \xbs' < \xbt' \xbt $ etc.

A comparison as in Example \ref{Example Seq} (page \pageref{Example Seq})
might also permit to compare sequences of different lengths.
\subsection{
Probability theory on partial orders using height
}

\label{Section Proba}

We define two notions of size of a set here:
 \xEh
 \xDH the size of a set is the maximal height of its elements, in
Section \ref{Section P} (page \pageref{Section P}), and
 \xDH the size of a set is the sum of the heights of its elements, in
Section \ref{Section P'} (page \pageref{Section P'}).
 \xEj

The first can be seen as a ``quick and dirty'' approach, the second
as a more standard one.
\subsubsection{
Size of a set as maximal height of its elements
}

\label{Section P}

\bd

$\hspace{0.01em}$


\label{Definition P}

 \xEh
 \xDH
For $X \xcc \xdx,$ we set $ht(X):=max\{ht(x):x \xbe X\}.$

If we are interested in $sup(X),$ we might define $ht(sup(X)):=ht(X)+1,$
and

if we are interested in $inf(X),$ we might define $ht(inf(X))$ $:=$
$min\{ht(x):x \xbe X\}-1.$
 \xDH
We may define a relative height by $rht(x):= \frac{ht(x)}{ht(\xct)},$
and we have $0 \xck rht(x) \xck 1,$ which may be interpreted as the
probability of $x.$

Thus, we define $P(x):=rht(x),$ and $P(X)$ similarly for $X \xcc \xdx.$
 \xEj

\ed

\bfa

$\hspace{0.01em}$


\label{Fact Schnitt}

We have the following facts for the height for $ \xew $ and $ \xfw:$
 \xEh
 \xDH
$ht(X \xew X')$ $<$ $ht(X),$ $ht(X'),$
 \xDH
$ht(X),$ $ht(X')$ $<$ $ht(X \xfw X')$
 \xEj

\efa

\subparagraph{
Proof
}

$\hspace{0.01em}$


This is trivial, as any chain to $(X \xew X')$ may be continued to a
chain to $X$ and
$X'.$ The second property is shown analogously.
Alternatively, we may use Fact 
\ref{Fact Height} (page 
\pageref{Fact Height}), (3).

$ \xcz $
\\[3ex]

\br

$\hspace{0.01em}$


\label{Remark Powerset}

When we work with subsets of some powerset, we use, unless defined
otherwise,
$ \xcB $ for $<,$ $ \xcs $ for $ \xew',$ $ \xcv $ for $ \xfw',$ and $
\xDN' $ is set complement.

\er

\be

$\hspace{0.01em}$


\label{Example Schnitt}

These examples show that $ht(X),$ $ht(X')$ may be arbitrarily bigger than
$ht(X \xew X'),$
and $ht(X \xfw X')$ may be arbitrarily bigger than $ht(X)$ and $ht(X').$

 \xEh
 \xDH
Let $ \xdx $ $:=$ $\{ \xCQ,$ $X_{1},$ $X_{2},$ $X'_{1},$ $X'_{2},$ $X,$
$X',$ $X \xcs X',$ $X \xcv X' \},$ with

$X:=\{a,a',b\},$ $X':=\{b,c,c' \},$ $X_{1}:=\{a\},$ $X_{2}:=\{a,a' \},$
$X'_{1}:=\{c\},$ $X'_{2}:=\{c,c' \}.$
Thus, $X \xcs X' =\{b\},$ and $ht(X \xcs X')=1,$ but $ht(X)=ht(X')=3.$

 \xDH
Let $ \xdx $ $:=$ $\{ \xCQ,$ $X_{1},$ $X_{2},$ $X,$ $X',$ $X \xcv X'
\},$ with

$X:=\{a,a' \},$ $X':=\{b,b' \},$ $X_{1}:=\{a,b\},$ $X_{2}:=\{a,a',b\}.$
Thus, $ht(X \xcv X')=3,$ but $ht(X)=ht(X')=1.$
 \xEj

$ \xcz $
\\[3ex]

\ee

\be

$\hspace{0.01em}$


\label{Example =1}

Some further examples:

 \xEh
 \xDH
$P(x)+P(\xDN x):$

$P(x)+P(\xDN x)$ may be 1, but also $ \xCc 1$ or $ \xCe 1:$
 \xEh
 \xDH
$=1:$

Consider $ \xdx $ $:=$ $\{ \xcT,a,b, \xct \}.$

Then $ \xDN a=\{b, \xcT \},$ $ \xDN' a=\{b\},$ and $P(a)=P(b)=1/2.$
 \xDH
$<1:$

Consider $ \xdx $ $:=$ $\{ \xCQ,$ $\{a\},$ $\{b,d\},$ $\{a,b,c\},$
$\{a,b,c,d\}\}.$

Then $ \xDN \{a\}$ $=$ $\{ \xCQ,\{b,d\}\},$ $ \xDN' \{a\}=\{\{b,d\}\}.$
Thus $ht(\{a\})=ht(\xDN' \{a\})=1,$ $ht\{a,b,c,d\}=3,$ and $P(\{a\})=P(
\xDN' \{a\})1/3,$ and
$1/3+1/3<1.$

 \xDH
$>1:$

Consider $ \xdx $ $:=$ $\{ \xCQ,$ $\{a\},$ $\{a,a' \},$ $\{b\},$ $\{b.b'
\},$ $\{a,a',b,b' \}\}.$

Then $ \xDN' \{\{a,a' \}\}$ $=$ $\{\{b,b' \}\},$ and
$ht\{a,a' \}$ $=$ $ht\{b,b' \}$ $=$ 2, $ht\{a,a',b,b' \}=3,$
so $P(\{a,a' \})=P(\xDN' \{a,a' \})=2/3,$ $2/3+2/3>1.$
 \xEj

 \xDH
$P(A \xfw B)$ may be $=,$ or $ \xCc,$ or $ \xCe $ $P(A)$ $+$ $P(B)$ -
$P(A \xew B).$

This follows from the above example, as $P(A \xew B)=0$ there.

 \xEj

\ee

We now consider independence, which we may define as usual:

\bd

$\hspace{0.01em}$


\label{Definition Ind1}

A and $B$ are independent iff $P(A \xew B)$ $=$ $P(A)*P(B).$

\ed

This, however, might be too restrictive, alternatives come to mind, e.g.

\bd

$\hspace{0.01em}$


\label{Definition Ind2}

A and $B$ are independent iff

$P(B):= \xba $ and

$P(A \xew B)/P(A)= \xba,$

or

$P(A \xew B)/P(A)< \xba,$
and $P((\xDN A) \xew B)/P(\xDN A) \xcg \xba,$

or

$P(A \xew B)/P(A)> \xba $ and
$P((\xDN A) \xew B)/P(\xDN A) \xck \xba,$

\ed

The best definition might also be domain dependent.
\subsubsection{
An alternative definition of P using height
}

\label{Section P'}

We turn to a more standard definition of size, based on point measure,
where the size of a point $x$ is again $ht(x),$ but the size of a set is
now the sum of the sizes of its points.

\bd

$\hspace{0.01em}$


\label{Definition Mu}

Alternatively, we may define for $X \xcc \xdx:$
 \xEh
 \xDH
$ \xbm (X)$ $:=$ $ \xbS \{ht(x):x \xbe X\},$
 \xDH
$P(X)$ $:=$ $ \frac{ \xbm (X)}{ \xbm (\xdx)}$
 \xEj

\ed

But we have similar problems as above with this definition, e.g. with
$P(x)$ and
$P(\xDN x),$ etc.:

If $x= \xcT $ or $x= \xct,$ then $P(x)+P(\xDN x)=1,$ but if
$x \xEd \xcT,$ and $x \xEd \xct,$ then $P(x)+P(\xDN x)<1,$ as $ \xct $
is missing.

Similarly, as $P(x \xew \xDN x)=0,$ we will often have $P(x \xfw \xDN x)$
$ \xEd $
$P(x)+P(\xDN x)-P(x \xew \xDN x).$

This, however is not due to incompleteness, as we can easily see by
considering complete partial orders. Consider e.g.
$ \xdx:=\{ \xcT,a,a',b,b', \xct \}$ with $a<a',$ $b<b'.$
Then $ \xDN a' =\{b,b' \},$ and $ht(a')=ht(b')=2,$ $ht(a)=ht(b)=1,$ $ht(
\xct)=3.$
$P(a')=2/9,$ $P(\xDN a')=3/9,$ $P(a' \xew \xDN a')=0,$ $P(a' \xfw \xDN
a')=3/9.$

Of course, for this definition of $P,$ considering a disjoint cover of
$ \xdx $ will have the desired property.
%
%
%
%
%
%
%
%
%
%
%
%
%
%
%
%
%
%
%
%

$ \xCO $

\end{document}